\documentclass[
preprint,
 amsmath,amssymb,
 aps,
pra
]{revtex4-2}
\usepackage{silence}
\usepackage{
    amsfonts,
    amsmath,
    amssymb,
    amsthm,
    bm,
    braket,
    dsfont,
    graphicx,
    latexsym,
    mathpazo,
    mathrsfs
}

\usepackage[hidelinks,pagebackref=true]{hyperref}

\usepackage{bm}

\usepackage{tikz}
\usetikzlibrary{quantikz2}

\usepackage{amssymb,amscd}
\usepackage{physics}
\usepackage{graphicx}
\usepackage{braket}
\usepackage{dsfont}
\usepackage{color}
\usepackage{tikz}
\usetikzlibrary{arrows,positioning}
\usetikzlibrary{shapes.geometric}
\usetikzlibrary{patterns}
\usepackage{mathrsfs}
\usepackage{mathtools}
\usepackage[utf8]{inputenc}
\usepackage{enumerate}
\usepackage[english]{babel}
\usepackage{lipsum}

\usepackage{appendix}

\newcommand{\gil}{\mathcal{\hada}}
\newcommand{\dens}{\mathcal{D}}
\newcommand{\bound}{\mathcal{B}}
\newcommand{\design}{\mathsf{C}_2}

\newcommand{\idmap}{\mathcal{I}}
\newcommand{\idmat}{I}
\newcommand{\myi}{\mathrm{i}}
\newcommand{\eref}{\eqref}
\newcommand{\sref}{\ref}
\newcommand{\fref}[1]{Figure~\ref{#1}}
\newcommand{\hada}{H}
\newcommand{\fidelity}{F}

\usepackage{nicefrac}

\usepackage{bm}

\numberwithin{equation}{section}

\DeclareMathOperator*{\average}{\mathbb{E}}

\newcommand{\tgate}{G}

\newcommand{\qudim}{d}
\newcommand{\auxt}{T}
\newcommand{\uniset}{\mathbf{U}}
\newcommand{\unimat}{U}
\newcommand{\uniop}{\hat{U}}

\newcommand{\matrep}{\Gamma}

\newcommand{\noise}{\mathcal{E}}

\newcommand{\twirlmatw}{W}
\newcommand{\twirl}{\mathcal{T}}

\newcommand{\channel}{\noise}

\usepackage{xspace}

\usepackage{amsthm}

\usepackage{thmtools}
\usepackage{thm-restate}
\declaretheorem[name=Theorem,numberwithin=section]{theorem}
\newtheorem{lemma}[theorem]{Lemma}

\theoremstyle{definition}

\usepackage{multirow}

\usepackage[draft,inline,nomargin]{fixme}
\fxsetup{theme=color}
\fxsetup{status=final}
\FXRegisterAuthor{da}{ada}{\color{purple}DA}

\usepackage[inline]{enumitem}

\makeatletter
\renewcommand*{\backref}[1]{}
\renewcommand*{\backrefalt}[4]{}
\makeatother

\begin{document}
\title{Characterisation of individual gates using twirling circuits}
\author{David Amaro-Alcal\'a\footnote{Corresponding author.}}

\affiliation{Research Centre for Quantum Information, Institute of Physics, Slovak Academy of Sciences, D\'ubravsk\'a cesta 9, Bratislava 845 11, Slovakia}
\affiliation{Institute for Quantum Science and Technology, University of Calgary,
 Alberta T2N~1N4, Canada}
 \affiliation{Department of Physics, Lakehead University, Thunder Bay, ON, P7B 5E1}

\begin{abstract}
  We present a method to characterise qubit gates. Utilising the supermap
  formalism, we create a scheme for deterministic single-qubit gate analysis.
  Our approach introduces a new twirling process that is applied directly
  through fixed circuits. This method removes the requirement to average over
  random gates. The results enhance randomised benchmarking techniques and are
  suitable for experimental setups with multi-qubit control, focusing on the
  precise characterisation of single-qubit gates.
\end{abstract}


\maketitle
\section{Introduction}

Supermaps play a central role in quantum information, driving applications in
resource theories and entanglement distillation~\cite{Wang2019,Gour2021}, and
more recently
enabling the representation of dephasing noise on gates, which is impossible to
simulate employing composition by channels~\cite{puchala2021,salazar2024}, and
experimentally tested~\cite{Li2024}.
Chiribella \emph{et al.}~\cite{Chiribella_DAriano_Perinotti_2008} presented the
most comprehensive realisation of
supermaps, which uses two isometry transformations on an extended system,
followed by a partial trace. Formally, supermaps define linear, completely
positive transformations between channels—CPTP maps acting on quantum states—and
appear under various names in the literature~\cite{Milz2021}.

This manuscript presents a family of unitary matrices implementing a quantum
channel transformation via the supermap formalism:
the transformation is a
twirling operation~\cite{Werner1989}.
Briefly, a twirling operation is the average (most of the time finite,
but it can use a non-uniform probability) over a group element of the
conjugation of an operator by the representation of a group acting on the same
Hilbert space.
The resulting channel transformation is
equivalent to twirling by the unitary group~\(\unimat(2)\).
This construction requires qudit systems,
which
are supported by
many platforms,
including ion traps and
superconductors, for preparation, evolution, and measurement.
Experimentally, our scheme requires preparing a qubit, a qutrit, and a ququart, or
four qubits.

We then apply our circuit to
estimate the average gate fidelity of a noisy target gate.
The scheme we introduce offers an alternative to interleaved
benchmarking (IB)~\cite{harper2017},
which is a variation of the randomised benchmarking (RB) scheme~\cite{Megesan2011}.
Our novel scheme requires that the circuit we propose remains unaffected by the target gate.
Recent studies of quantum volume indicate that
wider shallow circuits
are preferred over
deeper ones~\cite{Jurcevic2021, Bluvstein2023}.
Therefore, our scheme, which requires four extra qubits, 
is a viable alternative to IB.

The structure of the paper is outlined as follows: In the background section, we
review the concepts of supermaps and twirling operations, as well as the group
employed for the twirl.  In the approach section, we introduce the lemmata and
the theorem that form the core of this work, illustrating how to compute the
twirling process using the supermap formalism.  In the results section, we
present an explicit expression of the resulting twirling circuit and subsequently describe our scheme for estimating
the average gate fidelity of a channel.  After completing a draft of this work,
we became aware that a related preprint has appeared, which independently
develops a similar supermap formalism~\cite{Das2024}.
In that work, a twirl (although not named that) with respect to the
Pauli gate set is introduced. Their twirling method is used for different
purposes than gate characterisation.

\section{Background}

In this section, we outline key concepts, methods,
and established results necessary to
understand the material presented later.
The first subsection covers
twirling, including the technique and its requirements. 
Next, we discuss the
supermap formalism,
emphasising the motivation and the expressions that will be
used later in this paper. 
Lastly, in the third subsection,
we explain the IB scheme.


We begin by introducing the matrix representation of quantum channels,
which is needed to explain twirling.
Let \(\gil_\qudim\) be the Hilbert space of \(\qudim\)-level states
with basis \(\{\ket{0},\ldots, \ket{\qudim-1}\}\),
where \(\qudim\) is a positive integer.
These systems are called
qudits~\cite{brylinski2002universal}.
Let \(\bound(\gil_\qudim)\) be the set of bounded operators on~\(\gil_\qudim\).
Let \(\dens(\gil_\qudim)\) denote the subset of~\(\bound(\gil_\qudim)\)
corresponding to density operators;
the boundary of \(\dens(\gil_\qudim)\)
is the set of pure states.
A CPTP mapping from $\dens(\gil_\qudim)$ to~$\dens(\gil_\qudim)$
is denoted by~$\noise$~\cite{HeinosaariTeiko2012Tmlo}.

To introduce the matrix representation for qubit channels,
two operations are needed:~\(X\) and \(Z\),
which are defined as follows
\begin{equation}\label{eq:ops-x-z}
  X\ket{i} = \ket{i\oplus 1} \text{ and }
  Z\ket{i} = (-1)^i \ket{i}.
\end{equation}
From~\eref{eq:ops-x-z},
we denote the list of Pauli elements as
\begin{equation}\label{eq:pauli_elements}
	P_0 \coloneqq I,
	P_1 \coloneqq X, 
	P_2 \coloneqq Z,
	\text{ and }
	P_3 \coloneqq XZ.
\end{equation}
Now, the matrix representation for a channel $\noise$ can be defined as
the matrix $\matrep(\noise)$ with entries
\begin{equation}\label{eq:matrix-representation}
	\matrep(\noise)_{i,j} \coloneqq
	\tr[P_i^\dagger \noise(P_j)]/\sqrt{2}.
\end{equation}
For convenience,
given a unitary $\unimat$,
we denote the matrix representation of
the mapping~$\varrho\mapsto \unimat \varrho \unimat^\dagger$
as $\matrep(\unimat)$.
Note that the generalisation of this discussion
on channels and their representation can be extended
to qudits by using the qudit version of the Pauli 
matrices~\cite{Patera_Zassenhaus_1988}.


The notion of a twirl is connected to that of a group. The idea was first
introduced in quantum information by Ref.~\cite{Werner1989}, where it was used
to prepare a state invariant under any tensor product of two unitary matrices.
Later, in Ref.~\cite{Emerson_Alicki_Zyczkowski_2005}, the same technique was
employed to estimate the average gate fidelity of a gate set implementing
arbitrary unitary operations, which is also how we use it in this work.

The twirl of a channel \(\channel\) over a finite group \(\uniset\) is defined as:
\begin{equation}\label{eq:twirl-def-operation}
\twirl_{\uniset}[\noise]
\coloneqq 
\average_U
  \uniop\circ
	\noise\circ
	\uniop^\dagger,
\end{equation}
where \(\circ\) denotes composition of maps.
For any finite set of unitary matrices $\uniset$
with elements denoted by $\unimat$,
the matrix representation of the twirl by $\uniset$ is
\begin{equation}\label{eq:twirl_definition}
	\matrep(\twirl_{\uniset}[\noise])
	=
	\average_{\unimat\in \uniset}
  \matrep(\unimat)
	\matrep(\noise)
	\matrep(\unimat^\dagger),
\end{equation}
where 
$\average_{\unimat\in \uniset}$ denotes the uniform average
over $\uniset$.
In Sec.~\ref{subsec:rb},
we discuss the importance of
twirling in RB schemes,
in particular,
IB.

Alongside the Pauli gate set,
the following gate set is also appropriate.
We define
\begin{equation}\label{eq:the_group}
  \design \coloneqq\langle \auxt^t P: t \in \mathbb{Z}_3, P \in\{I, X, XZ, Z\}\rangle,
\end{equation}
with
\begin{equation}\label{eq:t-gate}
  \auxt\coloneqq
  \begin{bmatrix}
      1 & -\myi \\
      1 & \myi
  \end{bmatrix}/\sqrt{2},
  \quad\myi\coloneqq\sqrt{-1}.
\end{equation}
The group  $\design$
is a unitary 2-design~\cite{Wallman_2018}.
Therefore, twirling, a notion that is clarified in the following subsection,
with respect to $\design$,
is equivalent to twirling with respect to the unitary group.

  Important for reducing the number of gates needed to perform the twirl are
  non-simple groups.
  A simple group lacks non-trivial normal subgroups;
  a non-simple group then has at least one non-trivial normal subgroup.
  Let \(G\)
  be the whole group and \(N\) its normal non-trivial subgroup.
  Then we have the following result concerning the
  twirl with respect to \(G\): 
\begin{equation}
\twirl_{G}[\noise]
=
\twirl_{G/N}\circ\twirl_{N}[\noise]
,
\end{equation}
where \(\twirl_{G/N}\) denotes twirling over coset
representatives~\cite{Cross2016}.
  Thus, the twirl using G is the same as twirling
  only with respect to coset representatives and then twirling over the normal
  subgroup. As discussed in the following parts of the text, this leads to
  savings in the preparation of some gates.

Due to its importance in our work,
we describe the total depolarising channel.
The totally depolarising channel is a single-parameter, denoted by \(p\), quantum channel of the
form 
\begin{equation}\label{eq:depolarising}
  \mathcal{E}_{\text{dep}}(\rho; p)
  \coloneqq 
  p\rho + (1-p)\frac{\idmap}{\qudim} 
.
\end{equation}
The average gate fidelity of \(\mathcal{E}_{\text{dep}}(\rho; p)\) with respect to the
identity is given by
\begin{equation}\label{eq:link-fidelity-parameter-p}
\fidelity(\mathcal{E})
=
\frac{1+p(d-1)}{\qudim}
.
\end{equation}
From Eq.~\eqref{eq:link-fidelity-parameter-p}, we see that if the parameter \(p\) is known,
then the fidelity is known.  This fact is exploited in RB schemes, which is why
we are interested in mapping the noise channel to a depolarising channel.

The most general mapping between channels is
a supermap~\cite{Chiribella_DAriano_Perinotti_2008}.
The conditions 
imposed on supermaps are
linearity and complete positivity~\cite{Chiribella_DAriano_Perinotti_2008}:
linearity means that for any scalar \(k\) and two channels \(\noise_0\)
and \(\noise_1\), the action of a supermap \(\mathcal{S}\)
satisfies \(\mathcal{S}(\noise_0+k \noise_1) = 
\mathcal{S}(\noise_0)
+
k\mathcal{S}(\noise_1)
\);
the complete positivity refers to the fact that 
1) \(\mathcal{S}\) maps CP maps to CP maps
and 
2)
for any positive integer \(k\),
 \(\mathcal{S}\otimes \mathcal{I}_k\) maps 
CP maps to CP maps.

Considering these two conditions,
the supermap \(\mathcal{S}\) action on a channel \(\noise\) is
performed as
\begin{equation}
\mathcal{S}[\noise](\varrho)
=
\tr_{\mathcal{A}}\left[\twirlmatw
  \left(\noise \otimes \idmap_{\mathcal{B}}\right)\left(V \varrho V^{\dagger}\right)
\twirlmatw^{\dagger}\right]
=
\tr_{\mathcal{A}}[\hat{\twirlmatw}\circ 
 \left(\noise \otimes \idmap_{\mathcal{B}}\right)
 \circ\hat{V}(\varrho)]
,
\end{equation}
where $V$ is an isometry (not necessarily a unitary matrix)
with two auxiliary spaces~$\mathcal{A}$
and~$\mathcal{B}$
whose dimensions are related
to the number of Kraus operators of~\(\mathcal{S}\).
We note that in later parts of this work instead,
of labels such as
\(\mathcal{A}\),
we use an index for the subspace we trace.
In this paper, we construct the unitary matrices \(\twirlmatw\) and \(V\) that
result in a twirling operation over any channel \(\noise\) 
with the same effect as twirling over~\(\unimat(2)\),
or any unitary 2-design.

\subsection{Interleaved benchmarking}
\label{subsec:rb}

RB is widely regarded as the standard method for gate characterisation.
The main goal of most RB schemes
is to estimate the average gate fidelity of a target gate
or a target gate set (on average).
This performance metric provides a practical measure of how close an
implemented operation matches its ideal version.
By repeatedly applying random
gate sequences, these methods average out noise effects,
providing a reliable
estimate of gate quality.

Our proposed scheme provides an alternative to IB~\cite{Magesan2012,harper2017},
estimating the average gate fidelity of a single gate using a characterised
auxiliary gate set.
Such an approach can reduce the need for additional noise assumptions and
provide
greater flexibility in experimental settings.
The method is particularly useful in
scenarios where the auxiliary gates are already well understood,
allowing the focus to
remain on the gate of interest.

The noise of the target gate, the gate to be characterised,
is assumed to be: gate-independent, Markovian, 
and time-independent.
Under these assumptions,
the average gate fidelity of the target gate is the outcome
of the IB scheme~\cite{proctor2017}.
Whereas these constraints simplify the analysis,
they may not fully capture noise
in realistic devices~\cite{proctor2017,Brillant2025}.
Nevertheless,
they are widely used as a standard baseline in benchmarking literature.

We briefly discuss the IB scheme.
Note that only the figure of merit obtained in IB 
is relevant for our scheme;
the experimental processes are different.
In IB,  
a noisy gate \( \tgate \) is assumed to act as $\tgate \circ \noise$;
that is,
a composition of the ideal version of the gate \(\tgate\) with a channel
expected to be close to the identity map \(\noise\).
This formulation makes it possible to relate the fidelity of 
\(\tgate\) to that of \(\noise\) in a systematic way.

Therefore, estimating the average gate fidelity of the target gate is equivalent
to estimating the average gate fidelity of \(\noise\) with respect to the
identity map~\cite{Megesan2011}. By applying sequences of random operations with
some structure, one obtains a twirled version of \(\noise\). With respect to a
unitary 2-design, this takes the form
\begin{equation}\label{eq:twirl-depo}
\matrep(\twirl_{\unimat(2)}[\noise])
=
\matrep(\mathcal{E}_\text{dep})
\cong
\idmat_1 \oplus (1-p_{\mathcal{E}}) \idmat_3.
\end{equation} 
IB aims to estimate the parameter
\(p_{\mathcal{E}}\), which is related to the average gate fidelity of \(\noise\),
as described in Eq.~\eqref{eq:depolarising} and
Eq.~\eqref{eq:link-fidelity-parameter-p}. 

In this section, we introduced the state-of-the-art related to our work. In
particular, we review supermaps and IB. The supermap
formalism shows a way to design a circuit to perform any transformation for
channels, and the IB scheme estimates the fidelity of a
quantum gate. In the following section, we introduce our novel techniques for
performing a twirling operation and applying it to develop a scheme for
characterising gates.

\section{Approach}
\label{sec:approach}

In this section, we outline the key components of our results and their
interrelationships. The core concept is a supermap (and the corresponding
circuit), which involves averaging over \(\design\) group or any unitary
2-design. 
After discussing the details and assumptions, we use it to design a
scheme to characterise quantum gates and compare it with traditional RB schemes
that typically employ uniformly randomly sampled gates.


We start by discussing the construction of the circuit
and prove that its application results in the \(\design\)-twirl
of any channel.
We also present the corresponding circuit.
Then, aiming at our characterisation scheme,
we discuss the noise assumption.
The Hilbert space of a \(\qudim\)-level system is denoted by
\(\gil_\qudim\), which is spanned by pure states denoted by~\(\ket{i}\),
\(i\in[\qudim]\),
with
\begin{equation}
  [\qudim] \coloneqq  \{0,1,\ldots, \qudim-1\} .
\end{equation}
 This notation allows us to describe qudit systems in a compact and
 general way, covering both qubits and higher-dimensional qudits.

The density operators in \(\gil_\qudim\) are denoted by \(\varrho\),
and the CPTP mappings in the states \(\varrho\) are denoted by \(\noise\).
Unitary matrices acting on \(\ket{i}\) are denoted by \(\unimat\),
whereas
the associated unitary operator acting on \(\varrho\) is denoted by
\(\uniop\) as in \(\uniop(\varrho)\).
In the supermap formalism,
a mapping between channels is realised by two unitary operators
acting over an extended Hilbert
space.
Let \(\uniset = (U_0, \ldots,U_{d-1} )\) be an ordered finite set of \(\qudim\) unitary
 matrices over the Hilbert space of a
single qubit system.
Further assume the set \(\uniset\) forms a finite group.
This group structure plays a key role in simplifying the implementation of the twirl.

To obtain the twirl with respect to \(\uniset\),
we recall $\hada_d$, which is the Fourier transform matrix (Hadamard for \(\qudim=2\)):
\begin{equation}
  (\hada_\qudim)_{ij} \coloneqq 
  \exp(2\myi ij \pi/\qudim)/\sqrt{\qudim}.
\end{equation}
For convenience, we avoid writing the subscript \(\qudim\) as 
the dimension should be clear from the state on which \(\hada\) acts.
Consider the following unitary:
\begin{equation}
    \twirlmatw_{\uniset} \coloneqq  
    \sum_{i \in [d]} \unimat_i\otimes \hada\dyad{i}_\qudim \hada^\dagger.
\end{equation}
Then consider the composition of \(\hat{\twirlmatw}_{\uniset}\) with $\noise$
and tracing the second system:
\begin{equation}\label{eq:supermap}
\varrho\mapsto
    \tr_2
    [\hat{\twirlmatw}_{\uniset}\circ (\noise\otimes \mathcal{I}_\qudim)\circ
    \hat{\twirlmatw}_{\uniset}^\dagger(\varrho\otimes\dyad{0}_\qudim)]
    =
    \sum_{i \in [d]} \unimat_i^{\dagger}\noise(\unimat_i \varrho \unimat_i^{\dagger})\unimat_i;
\end{equation}
this is the twirl of \(\noise\) with respect to \(\uniset\).

\begin{lemma}[Twirling supermap]
Consider a set of \(\qudim\) unitary matrices \(\uniset\) acting on \(\gil_2\)
and a qubit channel \(\noise\).
The supermap performing the twirling operation
in Eq.~\eqref{eq:twirl-def-operation}
is given by
\begin{equation}\label{eq:definition-supermap}
  S[\noise](\varrho) \coloneqq 
  \tr_2
  [\hat{\twirlmatw}_{\uniset}\circ (\noise\otimes \mathcal{I}_\qudim)\circ
  \hat{\twirlmatw}_{\uniset}^\dagger(\varrho\otimes\dyad{0}_\qudim)],
\end{equation}
with 
\begin{equation}
    \twirlmatw_{\uniset}
    =
    \sum_{i \in [d]} \unimat_i\otimes \left(\hada\dyad{i}_\qudim \hada^\dagger\right).
\end{equation}
\end{lemma}

Composing two twirls may be more convenient than a single one when dealing
with groups that have a non-trivial normal subgroup. In that case, averaging
first over representatives of the coset and then over elements of the normal
subgroup is equivalent to averaging over the entire group. Thus, using the
normal subgroup and coset representatives achieves the same effect as applying
the procedure to the whole group but at a lower cost in terms of gate
preparation~\cite{Cross2016}. One example of a group with a non-trivial subgroup
is the gate set \(\design\) introduced in~\eref{eq:the_group}.

To attain the composition of twirls
in the supermap formalism,
we introduce the following lemma.
\begin{lemma}[Composition of twirls]\label{lemma:2}
Consider two sets of unitary matrices,
\(A = \{\unimat_i\}\) and~\(B = \{V_i\}\),
with \(A\) having \(\qudim_0\) matrices and \(B\) having \(\qudim_1\) matrices.
To obtain the following composition of two twirls:
\begin{equation}
\twirl_A\circ \twirl_B(\noise),
\end{equation}
the following unitary matrices are required
\begin{align}
  \twirlmatw_A &= \sum_{i\in [\qudim_0]} \unimat_i\otimes
  \left(\hada\dyad{i}_{\qudim_0}\hada^{\dagger}\right)\otimes \idmat_{\qudim_1}
  \text{ and}\label{eq:twirl_composition}\\
  \twirlmatw_B &= \sum_{i\in [\qudim_1]} V_i\otimes \idmat_{\qudim_0}\otimes
  \left(\hada\dyad{i}_{\qudim_1}\hada^{\dagger}\right).
\end{align}
Note that this makes the second auxiliary system a \(\qudim_0\)-level system
and the third auxiliary system a \(\qudim_1\)-level system.
Then form  
\begin{equation}\label{eq:mat_w}
  \twirlmatw_{A,B} \coloneqq  \twirlmatw_A \twirlmatw_B
\end{equation}
to get the twirl
\begin{equation}
\twirl_A\circ \twirl_B(\noise)
=
  \tr_{2,3}[\hat{\twirlmatw}_{A,B}\circ(\noise\otimes \idmap_{\qudim_0}\otimes
  \idmap_{\qudim_1}) \circ
  \hat{\twirlmatw}_{A,B}^{\dagger}].
\end{equation}
\end{lemma}
\begin{proof}
We present the proof in Appendix~\ref{app:proof}.
\end{proof}

In this section, we introduced the circuit to perform an operation on quantum
channels that is equivalent to twirling the channel with respect to a unitary
2-design. We explicitly constructed the matrices. In the following section, we
use the circuit constructed with these matrices to characterise a noisy gate by
estimating its average gate fidelity.
\section{Results}
In this section,
we apply Lemma~\ref{lemma:2}
to compute the
transformation equivalent to 
the
twirl with respect to the unitary group
and to introduce a scheme to characterise qubit gates.
We describe the complete procedure and give an explicit
bound for the number of experiments needed
using Hoeffding's inequality.
\subsection{Twirl}
\label{sub:twirl}

In this subsection,
using the results in \sref{sec:approach},
we show how to compute the \(\design\)-twirl.
Because the group 
\(\design\),
defined in Eq.~\eref{eq:the_group}, forms
a unitary 2-design,
the 
matrix representation, discussed in Eq.~\eqref{eq:matrix-representation}, of the twirl
of any channel \(\noise\) with respect to \(\design\) 
is 
\begin{equation}\label{eq:form_of_the_twirl}
    \matrep(\twirl_{\unimat(2)}[\noise])
    = \idmat_1\oplus (1-p(\noise)) \idmat_3.
\end{equation}
Next,
we show how to use Lemma~\ref{lemma:2}
to obtain the same transformation as twirling with respect to \(\design\)
but using a fixed circuit.

First, consider an auxiliary ququart
to compute the matrix
\begin{equation}
\unimat_P \coloneqq \sum_i P_i \otimes \hada\dyad{i}_4 \hada^{\dagger}\otimes \idmat_3,
\end{equation}
where 
\(P_i\) are Pauli operators introduced in~\eref{eq:pauli_elements}.
For the second part of the twirl,
the matrix is
\begin{equation}
\unimat_\auxt \coloneqq \sum_i \auxt^{i}\otimes \idmat_4\otimes \hada\dyad{i}_3 \hada^{\dagger},
\end{equation}
with \(\auxt\) defined in~\eref{eq:t-gate}.
Thus, the complete gate is
\begin{equation}\label{eq:twirling_gate}
\twirlmatw \coloneqq \unimat_P \unimat_\auxt.
\end{equation}
A simple computation using Lemma~\ref{lemma:2}
shows that using \(\twirlmatw\) in~\eref{eq:twirl_composition},
any input channel is transformed into its corresponding
twirl by \(\unimat(2)\).
This is our main contribution. We thus have shown that it is possible to
perform a twirl over any channel using a fixed circuit. Furthermore, by
exploiting a coset structure, the twirl requires fewer wires. This is relevant
to the many tasks~\cite{Elben2022} that use randomisation as a step in the
implementation of the protocol.
In the following section,
we use \(\twirlmatw\) in a supermap
to introduce a scheme to characterise noisy qubit gates.

\begin{figure}[h]
  \centering
  \includegraphics{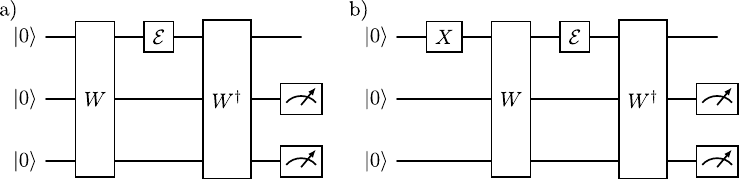}
\caption{\label{fig:circuit-primitive}
  a) Circuit implementing the \(\design\)-twirl.
  Time goes from left to right.
  The gate \(\twirlmatw\) is written in~\eref{eq:twirling_gate}.
  The measurement symbol amounts to discarding or tracing the state in that
  register.
  (From top to bottom) the first wire is a qubit,
  the second wire is a ququart,
  and the third wire is a qutrit.
  b) 
Circuit implementing the \(\design\)-twirl but now the initial state is
\(\ket{1}\).
 This step is necessary for the parameter estimation in 
Eqs.~\eqref{eq:system_of_equations}.
}
\end{figure}

\subsection{Characterisation scheme for individual gates}
In this subsection,
we discuss an application of our results in subsection~\ref{sub:twirl} to
characterise a single gate.
In contradistinction to IB,
our scheme does not require the gate-independent assumption;
that is, to assume that each gate has the same noise.
Consider a gate \(\tgate\) to be characterised,
with order \(k\); that is, \(\tgate^{k} = \idmat\).
The twirl of a noisy~\(\tgate^{k}\) applying a circuit in \fref{fig:circuit-primitive}
is denoted by \(\twirl_{\unimat(2)}[\tgate]\).

Recalling the form of the twirl in
Eq.~\eref{eq:form_of_the_twirl},
our scheme aims to estimate the parameter~\(p(\noise)\).
To obtain this parameter,
four
sequence fidelities 
are needed~\cite{Fogarty_Veldhorst_Harper_Yang_Bartlett_Flammia_Dzurak_2015}.
\begin{subequations}\label{eq:system_of_equations}
\begin{align}
  q_0 \coloneqq &\tr[\dyad{0}_2 \noise_0\circ \twirl_{\unimat(2)}[\tgate]\circ \noise_1(\dyad{0}_2) ],\\
  q_1 \coloneqq &\tr[\dyad{0}_2 \noise_0\circ \twirl_{\unimat(2)}[\tgate]\circ \noise_1(\dyad{1}_2) ],\\
  q_2 \coloneqq&\tr[\dyad{0}_2 \noise_0\circ
  \twirl_{\unimat(2)}[\idmap]\circ \noise_1(\dyad{0}_2) ],
  \text{ and}\\
  q_3 \coloneqq &\tr[\dyad{0}_2 \noise_0\circ
  \twirl_{\unimat(2)}[\idmap]\circ \noise_1(\dyad{1}_2) ].
\end{align}
\end{subequations}

The maps \(\noise_0\)
and \(\noise_1\)
represent the noise of  the supermap circuit.
The~\(X\) gate required is assumed to be noiseless.
Alternatively,
the noise can be included in the noise of the supermap circuit.

These \(q_i\) are obtained by four different circuits,
  which we proceed to explain:
  \begin{itemize}
    \item Use the circuit in Figure~\ref{fig:circuit-primitive}~a) to perform the twirl on \(\tgate^{k}\); that is, the
      target gate
      composed \(k\) times.
    \item Use the circuit in Figure~\ref{fig:circuit-primitive}~b) to perform the twirl on the target gate \(\tgate^{k}\);
      apply the \(X\) gate in every wire before measuring.
    \item Use the circuit in Figure~\ref{fig:circuit-primitive}~a) to implement the twirling without any gate, idle
      instead of \(\tgate^{k}\).
    \item Use the circuit Figure~\ref{fig:circuit-primitive}~b) to implement the twirling without any gate,
      but apply the \(X\) gate in every wire before measuring.
\end{itemize}
The first two correspond to using the circuit for twirling on \(\tgate^{k}\).  
For the last two, the twirling circuit is applied, but not \(\tgate^{k}\).  
Here we rely on the assumption that the noise of the twirling circuit is
independent of the implementation of \(\tgate^{k}\).  
This is the only assumption that our scheme—an alternative to
IB—requires.

Then, using~\eref{eq:system_of_equations},
we obtain
\begin{equation}
    1-p(\tgate) =
    \frac{q_0 - q_1}{q_2 - q_3}.
\end{equation}
In contradistinction to the RB scheme,
the number of experiments required to obtain a given confidence in the 
characterisation can be easily estimated
by using Hoeffding's
inequality~\cite{Hoeffding1963}.
Note that each sequence fidelity is the mean~\(q_i\),
written in Eqs.~\eqref{eq:system_of_equations},
of
a Bernoulli random variable (the distribution of
detection or lack of detection by the detector).
Thus, by applying Hoeffding's inequality, we obtain the following lower~bound on
the number of samples required:
\begin{equation}\label{eq:number-samples}
  N \geq
\frac{\log(2/\alpha)}{2\varepsilon^2},
\end{equation} where   \(\varepsilon\) is
the error and \(\alpha\) is the confidence.

Compared to RB sample complexity, which depends on
  circuit depths and experiments per depth, our scheme
  clearly states the necessary experiments. In RB, sample
  complexity remains an ongoing research topic due to
  fitting procedures and non-constant fidelity variance per
  depth~\cite{Wallman2014,helsen22prx}.  We include
  Eq.~\eqref{eq:number-samples} to highlight the fact that
our scheme has a precise sample estimation, a feature
currently unavailable for RB (including IB) schemes.

\danote*{Comment on applicability}{We conclude this subsection with a comment on the applicability of our
scheme.
For more general forms of noise, that is,
not only
\(\noise\circ\hat{\tgate}\), our scheme can be applied.
The only requisite is to be
able to justify that the noise of \(\hat{\tgate}^{k} = \idmap\) is the same as the single
operation of the gate. 
However, we foresee that the embedding of the hardware required for the gate
inside an existing circuit could pose experimental challenges.}
In this section,
we introduced our characterisation scheme,
showing how the fidelity can be estimated.
We also explicitly calculated the sample complexity.
In the following section, we will discuss these aspects and their implications.

\section{Discussion}
We constructed a circuit based on the supermap formalism that performs a
\(\design\)-twirl over any quantum channel. The circuit requires two auxiliary
qudits: one is a qutrit, and the other is a ququart. Based on the twirling
procedure, we designed a scheme to estimate the average gate fidelity of a gate.
Our scheme offers several advantages over IB, such as
avoiding the gate-independent assumption, and it is straightforward to provide
confidence intervals for the estimated parameter.

To our knowledge, this work is the first to utilise a supermap-based twirling
circuit in RB. By leveraging the coset structure of a unitary two-design, we
notably decrease the gate count. The circuit design builds upon earlier findings
related to dephasing channels~\cite{salazar2024}. Furthermore, using a fixed
circuit makes implementation easier than preparing random circuits on demand.
Additionally, our approach eliminates the need to fit an exponential decay,
leading to simpler and more reliable data analysis.

\section{Conclusion}

In this work, we constructed a supermap encoded in a circuit that maps an
arbitrary channel to a depolarising channel, maintaining the same fidelity as
the original.
This approach provides a foundation for developing devices that characterise
individual gates.
Our circuit design
reduces the number of required wires by exploiting the non-simplicity of
\(\design\);
From 16 wires (one per group element) to just three wires.
Furthermore, the transformation derived from the matrices
introduced in this work was applied to estimate the average gate fidelity of a
noisy gate.

The only assumption made is that the noise from the gate being characterised
does not impact the noise of the supermap circuit. This assumption is
potentially less restrictive than assuming gate-independent, Markovian, and
time-independent noise for the gate under consideration. By incorporating
supermaps into twirling operations, we have developed a novel characterisation
scheme that leverages the supermap formalism, thereby expanding the range of
benchmarking techniques available. Furthermore, this formalism could be adapted
for twirling operations on qudits or multi-qubit gates, utilising any of the
unitary two-designs for qudits while identifying an appropriate normal subgroup.
\section*{Acknowledgments}

We acknowledge support from the Natural Sciences
and
Engineering Research Council
of Canada, the Government of Alberta,
project
DeQHOST APVV-22-0570,
and project
QUAS VEGA 2/0164/25.
The author is grateful to Dr.\ Hubert de~Guise
and Dr.\ M\'ario Ziman
for comments on the manuscript and helpful discussions.
We also thank the anonymous referees who identified many issues in a previous version of this manuscript.
\bibliography{bibliography}

\providecommand{\noopsort}[1]{}\providecommand{\singleletter}[1]{#1}%
\begin{thebibliography}{10}

\bibitem{Bluvstein2023}
Bluvstein, D., S.~J. Evered, A.~A. Geim, S.~H. Li, H.~Zhou, T.~Manovitz, S.~Ebadi, M.~Cain, M.~Kalinowski, D.~Hangleiter, J.~P. Bonilla~Ataides, N.~Maskara, I.~Cong, X.~Gao, P.~Sales~Rodriguez, T.~Karolyshyn, G.~Semeghini, M.~J. Gullans, M.~Greiner, V.~Vuletić, and M.~D. Lukin.
\newblock \emph{Nature} \textbf{626} (2023)(7997), 58.

\bibitem{Brillant2025}
Brillant, A., P.~Groszkowski, A.~Seif, J.~Koch, and A.~A. Clerk.
\newblock \emph{Phys. Rev. Lett.} \textbf{135} (2025), 070601.

\bibitem{brylinski2002universal}
Brylinski, J.-L. and R.~Brylinski.
\newblock Universal quantum gates (2001).
\newblock ArXiv:quant-ph/0108062.

\bibitem{Chiribella_DAriano_Perinotti_2008}
Chiribella, G., G.~M. D'Ariano, and P.~Perinotti.
\newblock \emph{Europhys. Lett.} \textbf{83} (2008)(3), 30004.

\bibitem{Cross2016}
Cross, A.~W., E.~Magesan, L.~S. Bishop, J.~A. Smolin, and J.~M. Gambetta.
\newblock \emph{npj Quantum Inf.} \textbf{2} (2016)(1).

\bibitem{Das2024}
Das, S., J.~Sun, M.~Hanks, B.~Koczor, and M.~S. Kim.
\newblock Purification and correction of quantum channels by commutation-derived quantum filters (2024).
\newblock ArXiv:2407.20173.

\bibitem{Elben2022}
Elben, A., S.~T. Flammia, H.-Y. Huang, R.~Kueng, J.~Preskill, B.~Vermersch, and P.~Zoller.
\newblock \emph{Nat. Rev. Phys.} \textbf{5} (2022)(1), 9–24.

\bibitem{Emerson_Alicki_Zyczkowski_2005}
Emerson, J., R.~Alicki, and K.~\.{Z}yczkowski.
\newblock \emph{J. Opt. B: Quantum Semiclassical Opt.} \textbf{7} (2005)(10), S347.

\bibitem{Fogarty_Veldhorst_Harper_Yang_Bartlett_Flammia_Dzurak_2015}
Fogarty, M.~A., M.~Veldhorst, R.~Harper, et~al.
\newblock \emph{Phys. Rev. A} \textbf{92} (2015)(2), 022326.

\bibitem{Gour2021}
Gour, G. and C.~M. Scandolo.
\newblock \emph{Phys. Rev. A} \textbf{103} (2021), 062422.

\bibitem{harper2017}
Harper, R. and S.~T. Flammia.
\newblock \emph{Quantum Sci. Technol.} \textbf{2} (2017)(1), 015008.

\bibitem{HeinosaariTeiko2012Tmlo}
Heinosaari, T. and M.~Ziman.
\newblock \emph{{The Mathematical Language of Quantum Theory}}.
\newblock Cambridge UP (2012).

\bibitem{helsen22prx}
Helsen, J., I.~Roth, E.~Onorati, et~al.
\newblock \emph{PRX Quantum} \textbf{3} (2022), 020357.

\bibitem{Hoeffding1963}
Hoeffding, W.
\newblock \emph{J. Am. Stat. Assoc.} \textbf{58} (1963)(301), 13.

\bibitem{Jurcevic2021}
Jurcevic, P., A.~Javadi-Abhari, L.~S. Bishop, I.~Lauer, D.~F. Bogorin, M.~Brink, L.~Capelluto, O.~G\"{u}nl\"{u}k, T.~Itoko, N.~Kanazawa, A.~Kandala, G.~A. Keefe, K.~Krsulich, W.~Landers, E.~P. Lewandowski, D.~T. McClure, G.~Nannicini, A.~Narasgond, H.~M. Nayfeh, E.~Pritchett, M.~B. Rothwell, S.~Srinivasan, N.~Sundaresan, C.~Wang, K.~X. Wei, C.~J. Wood, J.-B. Yau, E.~J. Zhang, O.~E. Dial, J.~M. Chow, and J.~M. Gambetta.
\newblock \emph{Quantum Sci. Technol.} \textbf{6} (2021)(2), 025020.

\bibitem{Li2024}
Li, H., K.~Wang, S.~Wei, F.~Yang, X.~Chen, B.~C. Sanders, D.-S. Wang, and G.-L. Long.
\newblock \emph{New J. Phys.} \textbf{26} (2024)(1), 013037.

\bibitem{Megesan2011}
Magesan, E., J.~M. Gambetta, and J.~Emerson.
\newblock \emph{Phys. Rev. Lett.} \textbf{106} (2011), 180504.

\bibitem{Magesan2012}
Magesan, E., J.~M. Gambetta, B.~R. Johnson, C.~A. Ryan, J.~M. Chow, S.~T. Merkel, M.~P. da~Silva, G.~A. Keefe, M.~B. Rothwell, T.~A. Ohki, M.~B. Ketchen, and M.~Steffen.
\newblock \emph{Phys. Rev. Lett.} \textbf{109} (2012), 080505.

\bibitem{Milz2021}
Milz, S. and K.~Modi.
\newblock \emph{PRX Quantum} \textbf{2} (2021), 030201.

\bibitem{Patera_Zassenhaus_1988}
Patera, J. and H.~Zassenhaus.
\newblock \emph{J. Math. Phys.} \textbf{29} (1988)(3), 665.

\bibitem{proctor2017}
Proctor, T., K.~Rudinger, K.~Young, et~al.
\newblock \emph{Phys. Rev. Lett.} \textbf{119} (2017)(13), 130502.

\bibitem{puchala2021}
Pucha\l{}a, Z., K.~Korzekwa, R.~Salazar, P.~Horodecki, and K.~\ifmmode~\dot{Z}\else \.{Z}\fi{}yczkowski.
\newblock \emph{Phys. Rev. A} \textbf{104} (2021), 052611.

\bibitem{salazar2024}
Salazar, R. and F.~Shahbeigi.
\newblock \emph{Phys. Scr.} \textbf{100} (2025)(8), 085110.

\bibitem{Wallman_2018}
Wallman, J.~J.
\newblock \emph{Quantum} \textbf{2} (2018), 47.

\bibitem{Wallman2014}
Wallman, J.~J. and S.~T. Flammia.
\newblock \emph{New J. Phys.} \textbf{16} (2014)(10), 103032.

\bibitem{Wang2019}
Wang, X., M.~M. Wilde, and Y.~Su.
\newblock \emph{New J. Phys} \textbf{21} (2019)(10), 103002.

\bibitem{Werner1989}
Werner, R.~F.
\newblock \emph{Phys. Rev. A} \textbf{40} (1989), 4277.

\end{thebibliography}

\appendix
\section{Proof of Lemma~\ref{lemma:2}}\label{app:proof}

We start the proof of the double twirl. The result requires verifying that
extending, then conjugating, and then tracing leads to a twirl of an operation
\(\noise\).
Our proof starts by extending the state and conjugating by the matrix
\(\twirlmatw_{A,B}\).
For convenience, we denote by~\(\phi\) the density operator 
\begin{equation}
\phi_\qudim \coloneqq \dyad{0}_\qudim.
\end{equation}
We substitute the values of the matrices \(\twirlmatw_{A,B}\):

\begin{subequations}
\begin{align}
\twirlmatw_{A,B}^\dagger \, (\varrho \otimes \phi \otimes \phi) \, \twirlmatw_{A,B}
&= \twirlmatw_B^\dagger \twirlmatw_A^\dagger \, (\varrho \otimes \phi \otimes \phi) \, \twirlmatw_A \twirlmatw_B
\\&= \twirlmatw_B^\dagger \twirlmatw_A^\dagger \, (\varrho \otimes \phi \otimes \phi) 
\left( \sum_i U_i \otimes \hada |i\rangle \langle i| \hada^\dagger \otimes \idmat \right) \twirlmatw_B
\\&= \sum_i \twirlmatw_B^\dagger \twirlmatw_A^\dagger 
\left(\varrho U_i \otimes \phi  \hada |i\rangle \langle i| \hada^\dagger \otimes \phi \right) \twirlmatw_B
\\&= \sum_{i,i'} \twirlmatw_B^\dagger 
\left\{ U_{i'}^\dagger \otimes \hada |i'\rangle \langle i'| \hada^\dagger \otimes \idmat \right\}
\\&\qquad
\left(\varrho U_i \otimes \phi \hada |i\rangle \langle i| \hada^\dagger \otimes \phi \right) 
\twirlmatw_B
\\&= \sum_{i,i'} \twirlmatw_B^\dagger 
\left( U_{i'}^\dagger \varrho U_i \otimes \hada |i'\rangle \langle i'| \hada^\dagger \, \phi \, \hada |i\rangle \langle i| \hada^\dagger \otimes \phi \right) \twirlmatw_B
\\&= \sum_{i,i'} \twirlmatw_B^\dagger 
\Bigl( U_{i'}^\dagger \varrho U_i \otimes \hada |i'\rangle \langle i'| \hada^\dagger \, \phi \, \hada |i\rangle \langle i| \hada^\dagger
\otimes \phi\Bigr)
\\& \qquad 
 \Bigl(\sum_j V_j^\dagger \otimes \idmat \otimes \hada |j\rangle \langle j|
\hada^\dagger \Bigr) 
\\&= \sum_{i,i',j} \twirlmatw_B^\dagger 
\Bigl( U_{i'}^\dagger \varrho U_i V_j \otimes \hada |i'\rangle \langle i'| \hada^\dagger \, \phi \, \hada |i\rangle \langle i| \hada^\dagger
\\&\qquad\otimes \phi \, \hada |j\rangle \langle j| \hada^\dagger \Bigr)
\\&= \sum_{i,i',j,j'} 
\left( V_{j'}^\dagger \otimes \idmat \otimes \hada |j'\rangle \langle j'| \hada^\dagger \right)
\\&\qquad\left( U_{i'}^\dagger\varrho U_i  V_j \otimes \hada |i'\rangle \langle i'| \hada^\dagger \, \phi \, \hada |i\rangle \langle i| \hada^\dagger \otimes \phi \, \hada |j\rangle \langle j| \hada^\dagger \right)
\\&= \sum_{i,i',j,j'} 
\Bigl( V_{j'}^\dagger U_{i'}^\dagger\varrho U_i V_j
\\&\qquad\otimes \hada |i'\rangle \langle i'| \hada^\dagger \, \phi \, \hada
|i\rangle \langle i| \hada^\dagger \label{eq:wire1}
\\&\qquad\otimes \hada |j'\rangle \langle j'| \hada^\dagger \, \phi \, \hada
|j\rangle \langle j| \hada^\dagger \Bigr)\label{eq:wire2}
\end{align}
\end{subequations}


We now apply the channel \(\noise\) to the first wire and then conjugate by \(\twirlmatw_{A,B}\).

\begin{subequations}
  \begin{align}
&U_{k'} V_{l'} \noise \left( (U_i V_j)^{\dagger} \varrho U_i V_j \right) \left( U_k V_l \right)^{\dagger}
\\& \qquad\otimes \quad H \ket{k'} \bra{k'} H^\dagger H \ket{i'} \bra{i'} H^\dagger \phi H \ket{i} \bra{i} H^\dagger H \ket{k} \bra{k} H^\dagger
\\& \qquad\otimes \quad H \ket{l'} \bra{l'} H^\dagger H \ket{j'} \bra{j'} H^\dagger \phi H \ket{j} \bra{j} H^\dagger H \ket{l} \bra{l} H^\dagger\\
  &=U_{k'} V_{l'} \noise \left( (U_i V_j)^{\dagger} \varrho U_i V_j \right) \left( U_k V_l \right)^{\dagger}
\\& \qquad\otimes \quad H \ket{k'} \bra{k'}  \ket{i'} \bra{i'} H^\dagger \phi H \ket{i} \bra{i}  \ket{k} \bra{k} H^\dagger
\\& \qquad\otimes \quad H \ket{l'} \bra{l'}  \ket{j'} \bra{j'} H^\dagger \phi H \ket{j} \bra{j}  \ket{l} \bra{l} H^\dagger\\
  &=U_{k'} V_{l'} \noise \left( (U_i V_j)^{\dagger} \varrho U_i V_j \right) \left( U_k V_l \right)^{\dagger}
\\& \qquad\otimes \quad H \ket{k'} \delta_{k',i'} \bra{i'} H^\dagger \phi H \ket{i} \delta_{i,k} \bra{k} H^\dagger
\\& \qquad\otimes \quad H \ket{l'} \delta_{l',j'} \bra{j'} H^\dagger \phi H \ket{j} \delta_{j,l} \bra{l} H^\dagger\\
  &=U_{k'} V_{l'} \noise \left( (U_i V_j)^{\dagger} \varrho U_i V_j \right) \left( U_k V_l \right)^{\dagger}
\\& \qquad\otimes \quad\delta_{k',i'}\delta_{i,k} H \ket{k'}  \bra{i'} H^\dagger \phi H \ket{i}  \bra{k} H^\dagger
\\& \qquad\otimes \quad\delta_{l',j'}\delta_{j,l} H \ket{l'}  \bra{j'} H^\dagger \phi H \ket{j}  \bra{l} H^\dagger\\
  &=U_{i'} V_{j'} \noise \left( (U_i V_j)^{\dagger} \varrho U_i V_j \right) \left( U_i V_j \right)^{\dagger}
\\& \qquad\otimes \quad H \ket{i'}  \bra{i'} H^\dagger \phi H \ket{i}  \bra{i} H^\dagger
\\& \qquad\otimes \quad H \ket{j'}  \bra{j'} H^\dagger \phi H \ket{j}  \bra{j} H^\dagger\\
  \end{align}
\end{subequations}

Next, we trace over the second wire and then over the third wire.
\begin{subequations}
  \begin{align}
&\tr_{2,3} U_{i'} V_{j'} \noise \left( (U_i V_j)^{\dagger} \varrho U_i V_j \right) \left( U_i V_j \right)^{\dagger}
\\& \qquad\otimes \quad H \ket{i'}  \bra{i'} H^\dagger \phi H \ket{i}  \bra{i} H^\dagger
\\& \qquad\otimes \quad H \ket{j'}  \bra{j'} H^\dagger \phi H \ket{j}  \bra{j} H^\dagger\\
  &=U_{i'} V_{j'} \noise \left( (U_i V_j)^{\dagger} \varrho U_i V_j \right) \left( U_i V_j \right)^{\dagger}
\\& \qquad\otimes \quad \tr H \ket{i'}  \bra{i'} H^\dagger \phi H \ket{i}  \bra{i} H^\dagger
\\& \qquad\otimes \quad \tr H \ket{j'}  \bra{j'} H^\dagger \phi H \ket{j}  \bra{j} H^\dagger\\
  &=U_{i'} V_{j'} \noise \left( (U_i V_j)^{\dagger} \varrho U_i V_j \right) \left( U_i V_j \right)^{\dagger}
\\& \qquad\otimes \quad \tr  \ket{i'}  \bra{i'} H^\dagger \phi H \ket{i}  \bra{i} 
\\& \qquad\otimes \quad \tr  \ket{j'}  \bra{j'} H^\dagger \phi H \ket{j}  \bra{j} \\
  &=U_{i} V_{j} \noise \left( (U_i V_j)^{\dagger} \varrho U_i V_j \right) \left( U_i V_j \right)^{\dagger}
\\& \qquad\otimes \quad  \bra{i} H^\dagger \phi H \ket{i} 
\\& \qquad\otimes \quad  \bra{j} H^\dagger \phi H \ket{j} \\
  &= \frac{1}{\qudim^2}
  U_{i} V_{j} \noise \left( (U_i V_j)^{\dagger} \varrho U_i V_j \right) \left( U_i V_j \right)^{\dagger}.
  \end{align}
\end{subequations}

Summing over \(i\) and then over \(j\)  we obtain:
\begin{equation}
\frac{1}{\qudim^2}
\sum_{i,j}
  U_{i} V_{j} \noise \left( (U_i V_j)^{\dagger} \varrho U_i V_j \right) \left( U_i V_j \right)^{\dagger}
  =
\twirl_A\circ \twirl_B(\noise).
\end{equation}
Which is the expression we wanted to prove.

\end{document}